\documentclass{aa}  
\usepackage{graphicx}
\usepackage{orcidlink}
\usepackage{txfonts}
\usepackage{hyperref}

\usepackage{xcolor}
\usepackage{soul}

\begin{document} 
   \title{Wave transformations near a coronal magnetic null-point}

%   \subtitle{I. Overviewing the $\kappa$-mechanism}
  \author{
  N. Yadav
  \inst{1,2}\orcidlink{0000-0002-8976-516X}
  \and
  Rony Keppens\inst{1}\orcidlink{0000-0003-3544-2733}
}
  % \institute{Department of Physics, Indian Institute of Science Education and Research Thiruvananthapuram, 695551 Kerala, India
  % \and
  % Centre for mathematical Plasma Astrophysics, Department of Mathematics, KU Leuven, Celestijnenlaan 200B, 3001 Leuven, Belgium  %, %\institute{
  \institute{Centre for mathematical Plasma Astrophysics, Department of Mathematics, KU Leuven, Celestijnenlaan 200B, 3001 Leuven, Belgium  %, %\institute{
  \and
  Department of Physics, Indian Institute of Science Education and Research Thiruvananthapuram, 695551 Kerala, India
  \\
                \email{nitnyadv@gmail.com}
%          \and
% School of Physics and Astronomy, Yunnan University, 650050 Kunming, People’s Republic of China
            }
   \date{}
% \abstract{}{}{}{}{} 
% 5 {} token are mandatory
 
  \abstract
   {Null points are often invoked in studies of quasi-periodic coronal jets, and in connection with periodic signals preceding actual reconnection events.
   Although the periodicity of such events spans a large range of periods, most show 2-5 minute periodicity compatible with the global p-modes.}
   {We aim to investigate the viability of MHD waves, in particular acoustic p-modes, in causing strong current accumulation at the null points. This can in turn drive localized periodic heating in the solar corona.}
   {To reach this goal, we begin with a three-dimensional numerical setup incorporating a gravitationally stratified solar atmosphere and an axially symmetric magnetic field including a coronal magnetic null point. To excite waves, we employ wave drivers mimicking global p-modes.
   Using our recently developed wave mode decomposition technique, we investigate the process of mode conversion, mode transmission, and wave reflection at various important layers of the solar atmosphere, such as the Alfv\'en acoustic equipartition layer and transition region.
   We examine the energy flux distribution among various MHD modes or among acoustic and magnetic components, as the waves propagate and interact with a magnetic field of null topology. We also examine current accumulation in the surroundings of the null point.}
   {We found that most of the vertical velocity transmits through the Alfv\'en acoustic equipartition layer maintaining acoustic nature while a small fraction generates fast waves via the mode conversion process. The fast waves undergo almost total reflection at the transition region due to sharp gradients in density and Alfv\'en speed. There are only weak signatures of Alfv\'en wave generation near the transition region due to fast-to-Alfv\'en mode conversion. Since the slow waves propagate with the local sound speed, they are not much affected by the density gradients at the transition region and undergo secondary mode conversion and transmission at the Alfv\'en-acoustic equipartition layer surrounding the null point, leading to fast wave focusing at the null point. These fast waves have associated perturbations in current density, showing oscillatory signatures compatible with the second harmonic of the driving frequency which could result in resistive heating and enhanced intensity in the presence of finite resistivity.}
   {We conclude that MHD waves could be a potential source for oscillatory current dissipation around the magnetic null point. 
   We conjecture that besides oscillatory magnetic reconnection, global p-modes could lead to the formation of various quasi-periodic energetic events.}


   \maketitle
%
%-------------------------------------------------------------------

\section{Introduction}
 %%%%%%%%%%%%%%%%%%%%%%%%%%%%%%%%%%%%%%  
Granular buffeting and turbulent plasma motions on the solar surface excite a wide range of waves propagating through the solar atmosphere, which is pervaded by magnetic fields (\citealt{1981A&A....98..155S,1995A&A...304L...1V,2014A&A...569A.102S}).
The magnetic field configuration influences the propagation, mode conversion, and dissipation of these waves. These intricate wave-related processes deserve to be studied in idealized setups with specific, but representative magnetic field geometries. In an ideal MHD setting, the triplet of slow, Alfv\'en, and fast waves from uniform plasma settings allow for various linear wave transformations.

\cite{2006ApJ...653..739K} investigated the propagation of fast magnetoacoustic waves from the photosphere to the low chromosphere in an azimuthally symmetric flux tube resembling a magnetic sunspot.
They observed that depending on the frequency of the incoming fast (acoustic) wave and the angle between wave vector and ambient magnetic field, the wave undergoes partial conversion to a fast (magnetic) wave and partial transmission to slow (acoustic) wave at the Alfv\'en-acoustic equipartition layer, where sound and Alfv\'en speeds are equal (\citealt{2005MNRAS.358..353C,2006RSPTA.364..333C}).
The fast (magnetic) wave then travels up in the chromosphere and suffers from refraction and complete reflection due to a vertical gradient in phase speed. The 2.5D assumption made by \cite{2006ApJ...653..739K} precluded any fast-to-Alfv\'en mode conversions and suggested that fast waves are ineffective in heating the atmosphere as they are returned back to the photosphere. \cite{2021A&A...653A.131P} recently re-visited the fast-to-slow mode transmission in 2.5D settings, extended with ambipolar diffusion effects due to the partially ionized chromospheric regions. Rather short period fast waves, traveling across the (inclined) field were found to damp before they undergo reflection.

Fast-to-Alfv\'en mode conversion occurs at or above the turning height for fast waves, another key phenomenon of fast wave propagation through the solar atmosphere.
For a uniform inclined magnetic field, it is shown that this mode conversion is most efficient for angles $\theta$ between the magnetic field vector and the vertical lying between $30^{\circ}-40 ^{\circ}$, and for angles $\phi$ between the magnetic field vector and the wave propagation vector lying between $60^{\circ}-80^{\circ}$ (\citealt{2008SoPh..251..251C,2013JPhCS.440a2047M}). 
This conversion process drains energy from the reflecting fast wave and changes its phase before it returns.
Alfv\'en waves generated by this conversion process travel upwards/downwards depending on the ``field azimuth'', which is the magnetic field orientation to the vertical plane of wave propagation (\citealt{Khomenko_2012}).
Alfv\'en waves propagating upwards encounter steep density gradients at the transition region and get partly reflected (\citealt{2005ApJS..156..265C}).
However, a small fraction of Alfv\'en waves escape through the transition region and this has essential implications for coronal heating and solar wind acceleration (\citealt{2015NatCo...6.7813M}).

A null point in the coronal magnetic field configuration is also a common component in MHD models invoked to explain coronal jet formation. Coronal magnetic field is dominantly frozen into the plasma, however, strong currents are detected at the quasi-separatrix layers  and are reported to be responsible for initiating coronal jets (\citealt{2022FrASS...920183S}). Thus coronal null points favor the occurrence of jets.
In that context, nonlinear MHD simulations investigate the process of induced magnetic reconnection when such a magnetic topology is energized and often emphasize the null point role in coronal jet formations (\citealt{2009McLaughlin}). 
In the process of magnetic reconnection, free magnetic energy is converted into kinetic energy of plasma and non-thermal particles get accelerated (\citealt{2002A&ARv..10..313P}).
Though there are various observational and simulation studies suggesting magnetic reconnection to be causing quasi-periodic coronal jets, several coronal observations prove the presence of a magnetic null topology before coronal jet formation (\citealt{1992PASJ...44L.173S,2008ApJ...673L.211M,2013A&A...559A...1S}) indicating that magnetic reconnection need not necessarily be the source of quasi-periodic jets.
\citet{2020A&A...639A..22J} compared and discussed coronal jets emerging from an active region using observational data from AIA and IRIS. 
They showed that all six coronal jets in their study showed pre-jet intensity oscillations at their base with oscillation periods ranging between 1.5 to 6 min accompanied by smaller jets.
Similar periodic intensity oscillations have earlier been reported by \citealt{2018ApJ...855L..21B} for quiet-region jets.
Those periodic intensity variations, with periods in the several-minute range, may signal pre-jet conditions, which is yet another motivation for our present study.

In the present work, we aim to investigate the implications of mode conversion and mode transmission processes on current accumulation and heating in the vicinity of a magnetic null point.  To this end, we extend our previous ideal 3D magnetohydrodynamic (MHD) numerical study (\citealt{2022A&A...660A..21Y}) that put forward a new MHD wave decomposition method and investigated the interaction of waves generated by photospheric vortices with a coronal null, which had an axially symmetric magnetic field setup. 
Such a 3D null configuration is a common ingredient in actual solar atmospheric magnetic topologies and introduces a fan surface collecting field lines originating from the null, as well as a (vertical) spine. When a magnetic dipole is nested within a larger-scale domain of single polarity, a magnetic null point naturally emerges.
In our previous study, the applied bottom wave driver was a spiral driver; most of the energy was in the form of Alfv\'en modes that spread out the fan plane resulting in current localization at the separatrices (\citealt{2022A&A...660A..21Y}).
In agreement with earlier studies (\citealt{2003JGRA..108.1042G, 2004A&A...420.1129M}), we found that currents are accumulated at the fan plane that separates regions with different magnetic flux connectivity. The rotational wave driver was motivated by the omnipresent vortex flows, seen in both observations and simulations of  photospheric dynamics (\citealt{1988brandt,Wedemeyer-Bohm2012}). Another omnipresent ingredient at photospheric layers is the compressive p-modes, and here we study the interaction of global p-modes with a null point configuration with the aim of investigating their role in oscillatory heating processes often observed in the vicinity of magnetic null points.

We investigate the details of wave mode transformations at the Alfv\'en-acoustic equipartition surface surrounding a null point where sound speed ($c_s$) is equal to Alfv\'en speed ($v_A$). Inspired by actual low plasma beta coronal conditions, most of our simulation domain will have low beta values, implying negligible coupling between the three fundamental MHD modes.
However, modes get coupled near the Alfv\'en-acoustic equipartition layer which surrounds the coronal null region.
Numerous numerical studies investigated MHD wave propagation near nulls in great detail over the past decade (\citealt{2012ApJ...758...96F, 2015A&A...577A..70S,2017A&A...602A..43S,  2017ApJ...837...94T,  2019ApJ...879..127T}).
However, most of these works did not pay much attention to mode conversion surrounding the null point and excitation of fast-wave there.
\citet{2006A&A...459..641M} investigated the propagation of fast magneto-acoustic waves through a magnetic null point in a 2D geometry, assuming uniform background pressure and density.
They found that waves undergo mode conversion at the Alfv\'en-acoustic equipartition layer surrounding the null point. Their study inserted fast mode perturbations from the top boundary and neglected gravitational stratification and 3D effects. In the current study, we generalize this setup to a 3D magnetostatic null with photosphere-to-corona conditions, insert acoustic p-mode power at the bottom layers, and verify their conversion to fast (magnetic) waves in the corona, and their ultimate transformations/interactions near a null.

According to the polar diagram for the group velocities of fundamental MHD modes, slow waves roughly follow magnetic field lines while drifting somewhat from them. As a result, it seems unlikely that a slow wave can reach the magnetic null points. Whereas, fast waves can easily approach the null point because they can propagate across the magnetic field lines. Additionally, due to wave refraction, a fast magnetoacoustic wave is drawn to the magnetic null point (\citealt{2004A&A...420.1129M, 2009SSRv..149..119N}). 
Fast magnetoacoustic waves are vastly observed in the range of 1-4 minutes and they are usually found to be closely associated with flare energy releases (\citealt{2011ApJ...736L..13L,2012ApJ...753...52L,2016ApJ...822....7K}).
\citet{2017ApJ...844..149K} investigated the impulsive phase of an M-class flare and reported quasi-periodic rapidly propagating fast-mode waves with periods of 120 to 240 s and simultaneous quasi-periodic bursts with periods $\sim$ 70 and $\sim$ 140 s, which are compatible with the second harmonic of fast-wave periodicity. 
They showed that fast-wave signals are not detected before the flare but are observed after the flare has initiated at the null point and oscillatory signatures are detected in the SXR flux. 
Thus, although fast wave focusing at the null point could be a potential physical mechanism leading to the quasi-periodic enhancement in the intensity at the null point, their source of origin is yet not very clear.

Since fast waves could be generated by mode conversion at the Alfv\'en-acoustic equipartition layer (\citealt{2005MNRAS.358..353C,2006RSPTA.364..333C}), a fast-mode MHD periodic wave, essentially excited due to mode conversion of acoustic p-modes could be the source causing quasi-periodic energetic events. To examine the role of the ubiquitous p-modes in such periodic intensity enhancements, we investigate their interaction with a coronal magnetic null point.
In our 3D ideal MHD numerical experiment, we use a similar magnetic topology detected by \citet{2020A&A...639A..22J} and investigate the linear wave mode conversion, transmission, and reflection processes at various essential layers in the solar atmosphere. We investigate the role of acoustic p-modes in the generation of quasi-periodic heating at the null points.

The paper is organized as follows: we recall the numerical setup in Section \ref{numerical_setup}. We discuss the obtained results and their implications in Section \ref{results}.  We finally conclude in Section \ref{conclusion}.
% \newpage

\begin{figure*}
     \centering
        \includegraphics[scale=0.9,trim=0 0 0 0]{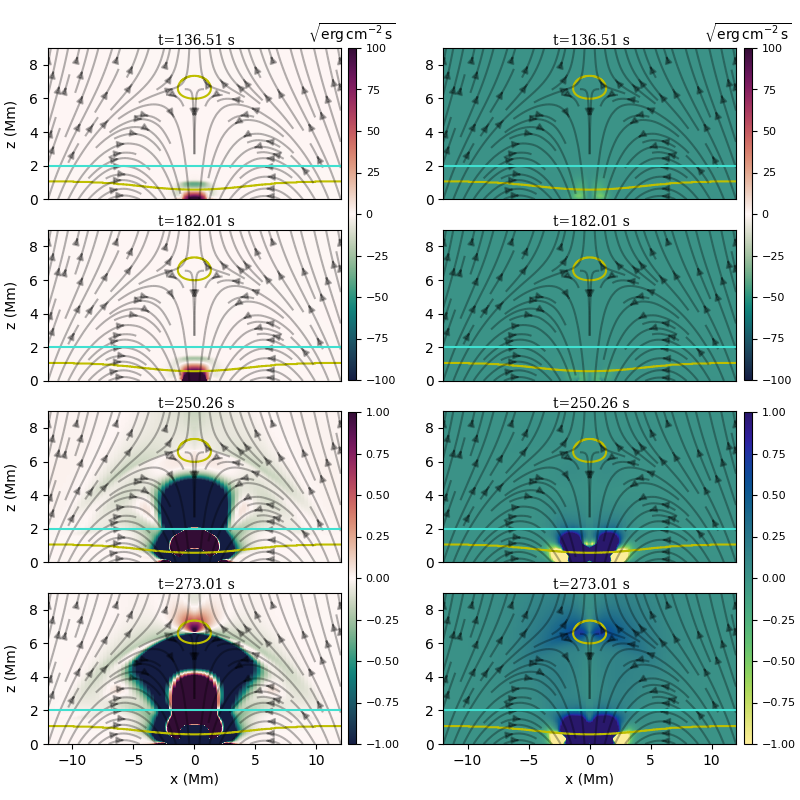}
             \caption{Slow and fast wave evolutions in a vertical cutting plane through the 3D null spine. Spatial maps of slow (left column) and fast (right column) wave amplitudes (in terms of their associated wave energy fluxes i.e. $\sqrt{\rho_0 c_s}v_{slow}$ and $\sqrt{\rho_0 v_A}v_{fast}$) in the $y=0$ plane at various time instants (time evolves from top to bottom, mentioned on top of each panel). Yellow curves represent the Alfv\'en-acoustic equipartition layers while the turquoise line displays the bottom of the transition layer at $z=2$ Mm. Maps are over-saturated to bring out dynamics taking place close to these crucial layers.}
             \label{mode_conversion}
\end{figure*}
\begin{figure*}
     \centering
 \includegraphics[scale=0.9,trim=0 0 0 0]{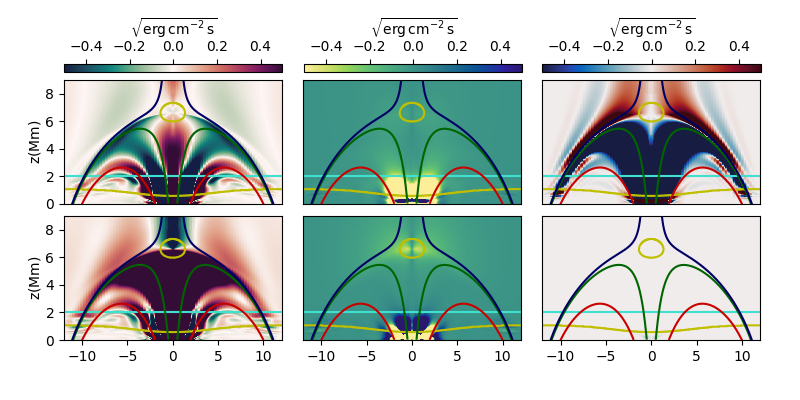}
             \caption{Comparing the wave transformations between rotational and p-mode drivers. Maps of slow wave (left column), fast wave (middle column), and Alfv\'en wave (right column) at one instant close to the end of our time series (i.e. $\sim$ 47 minutes for both the simulations). The top panel shows the distribution when applying a rotational driver at the bottom boundary, while the bottom panel corresponds to vertical velocity driving.}
             \label{one_instant_with_surface_lines}
\end{figure*}

\section{Numerical Setup}\label{numerical_setup}
We use the open-source \texttt{MPI-AMRVAC} \footnote{http://amrvac.org} (\citealt{2018ApJS..234...30X,KEPPENS2021316, 2023A&A...673A..66K}) code to perform simulations using the same numerical setup as in \citet{2022A&A...660A..21Y} except for the profile of the velocity driver imposed at the lower boundary for wave excitation.
We refer the readers to \citet{2022A&A...660A..21Y} for more details on the numerical setup and the constant magnetic flux surfaces selected for analysis. We recall that we are solving the full nonlinear ideal 3D MHD equations, where we solve for the deviations in pressure, density, and magnetic field as compared to a magnetostatic 3D stratified equilibrium that features an axisymmetric null configuration with a dome-spine structure. The stratification goes from the photosphere to the corona and has a temperature profile similar to a coronal transition region at 2-3 Mm height. In part of the later analysis, we will look at wave variations that can be quantified on (axisymmetric, nested) flux surfaces, where we differentiated a (red and green) flux surface within the dome, from a (blue) flux surface outside the dome.

In contrast to the previous work, we now modify the wave driver profile but still locate it purely underneath the central part of the dome (the dome has a roughly 12 Mm radius). We now apply a wave driver in the form of a vertical velocity perturbation that oscillates in time with a period of 4 minutes and extends in a circular region of a radius of 1 Mm.
It is chosen such that it represents global p-modes and has the form $v_z=v_0 \, \mathrm{sin}(\omega t)$
where $v_0$=20 m/s is the amplitude, $\omega$ is the oscillation frequency.
The motivation behind employing the p-mode driver in this region is two-fold. Firstly, we intend to make a direct comparison with our previous study and, therefore, employ the wave driver with the same spatial extent.
Secondly, p-modes are representative of convective phenomena that have typical spatial scales of 1-1.5 Mm.
It is important to note that acoustic waves with higher and lower frequencies will have, correspondingly, smaller and larger spatial scales (\citealt{2021A&A...652A..43Y}).
Moreover, we took a smaller amplitude of the driving velocity so that perturbations remain smaller compared to the background quantities even after the wave propagates through our gravitationally stratified atmosphere.
We do this to focus on the linear to the weakly nonlinear regime, so we can compare our results directly with our earlier study (\citealt{2022A&A...660A..21Y}). This weakly nonlinear regime is also relevant for the pre-coronal jet conditions, where similar periodicities were reported.
In comparison to our previous study (\citealt{2022A&A...660A..21Y}, we take a much smaller amplitude here such that the MHD wave amplitudes in the higher layer remain in a similar order of magnitude in both cases.

\section{Results and Discussion}\label{results}
We first discuss the overall evolution of the waves as they travel through our 3D stratified atmosphere, enriched by a magnetic null topology. Note that the central field underneath the dome has about 150 G strength near the photosphere, where the plasma beta is above unity. A small high beta region also surrounds our magnetic null, but most of the domain above 1 Mm height is in a low beta state. In the discussion below, we will make use of our novel, field-line-geometry-based wave mode decomposition introduced in our previous work (\citealt{2022A&A...660A..21Y}) to calculate the velocity components associated with the three fundamental MHD waves. This will allow us to uniquely quantify slow, Alfv\'en, and fast wave components, necessarily most accurate in the low beta regions. We will thereby also compare the results obtained with the rotational driver with the p-mode driver.

\subsection{Wave mode conversion, transmission, and reflection}

Fig.\ref{mode_conversion} displays various time instants that illustrate the mode conversion process near the Alfv\'en-acoustic equipartition layers and at the bottom of the transition region (i.e. around 2 Mm above the mean solar surface). Time evolves from the top row to the bottom row.
The left column displays the slow wave component (acoustic) scaled in terms of $\sqrt{\rho_0 c_s}v_{slow}$, while the right column displays the fast wave component (magnetic) scaled in terms of $\sqrt{\rho_0 v_A}v_{fast}$, of the perturbation traveling upwards that is essentially resulting from the vertical velocity perturbation imposed at the lower boundary representing global p-modes.
The left panel in the top row shows that the perturbation is almost parallel to the background magnetic field in the region of the applied driver and undergoes almost complete transmission through the equipartition layer with minimal mode conversion.
The right panel of the top row shows a weak signal resulting from mode conversion on the locations where the magnetic field is more inclined. 
At a slightly later stage, at t=182.01 s, the slow wave signal grows further, traveling with the local sound speed. However, the wave signal in the right column is negligible. 
The third row from the top corresponds to the snapshot at t=250.26 s. 
Here we see that the slow wave travels much faster above 2 Mm than below it, because the local sound speed increases after crossing the transition region. On the other hand, the fast wave undergoes almost complete reflection near that height due to the very sharp density and Alfv\'en speed gradients.
The last row corresponds to an instant when the upwards traveling slow waves encounter another Alfv\'en-acoustic equipartition layer that surrounds the null.
The slow waves then undergo  partial transmission and partial mode conversion depending on the attack angle, that is, the angle between the incoming wave vector and the background magnetic field. Since the background magnetic field is dominantly parallel to the incoming wave vector, there is very little mode conversion to a fast wave as seen in the right panel of the bottom row and most of the perturbation transmits through the Alfv\'en-acoustic equipartition layer without change in the acoustic nature of the wave.

To better appreciate the difference in the wave mode evolutions, we compare results with our earlier rotational driver with the current vertical driver setup.
Fig. \ref{one_instant_with_surface_lines} displays the scaled velocity perturbations associated with all three MHD waves, viz. slow wave, fast wave, and Alfv\'en wave in the left, middle, and right column, respectively, for the rotational driver (top row) and vertical driver (bottom row). With the rotational driver, most wave energy flux ended up in Alfv\'en waves, with a substantial fraction of energy flux in the slow wave. 
Fast waves are generated due to the mode conversion of slow waves at the Alfv\'en-acoustic equipartition layers. 
Fast waves reflect and are trapped below the transition region, with almost no accumulation of fast waves near the null point (\citealt{2022A&A...660A..21Y}).
In sharp contrast, for vertical driving, most flux is in the slow mode, with negligible flux in the Alfv\'en mode. 

\begin{figure*}
     \centering
        \includegraphics[scale=0.58]{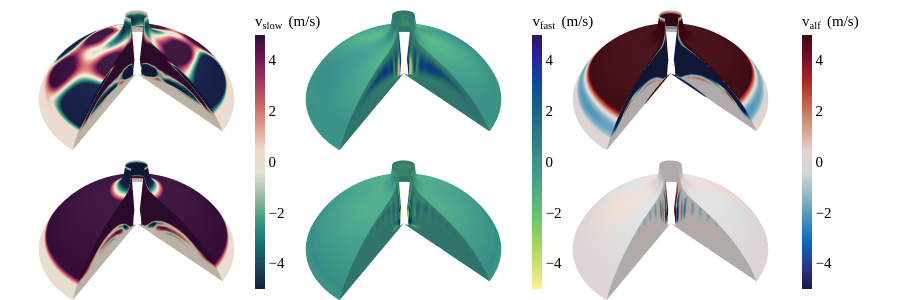}
             \caption{Velocity perturbations among slow, fast and Alfv\'en modes. Three-dimensional cross-sectional view at the end of our time series for slow (left), fast (middle), and Alfv\'en wave (right). The top and bottom rows correspond to rotational and vertical driving, respectively.}
             \label{3D_cross_section}
\end{figure*}
To visualize how the waves behave over the entire 3D region, we show the velocity perturbations in Fig. \ref{3D_cross_section} in a three-dimensional cross-sectional view of the domain. The upper panel corresponds to rotational driving, while the bottom row corresponds to vertical driving.
The results are shown for one instant; a movie for the time evolution during the full-time series is available as supplementary material.
A few results can be straightforwardly drawn from this figure. The bottom row confirms that the slow mode is dominant, while exciting waves of modest amplitudes in fast and Alfv\'en families with the vertical velocity driver. There is negligible power in the fast mode in both cases. Most of the wave power is in the Alfv\'en mode in the case of the rotational driver, while there is not much Alfv\'en power in the domain when using a vertical velocity driver.

\begin{figure*}
     \centering
        \includegraphics[scale=0.4,trim=0cm 0cm 0cm 0cm]{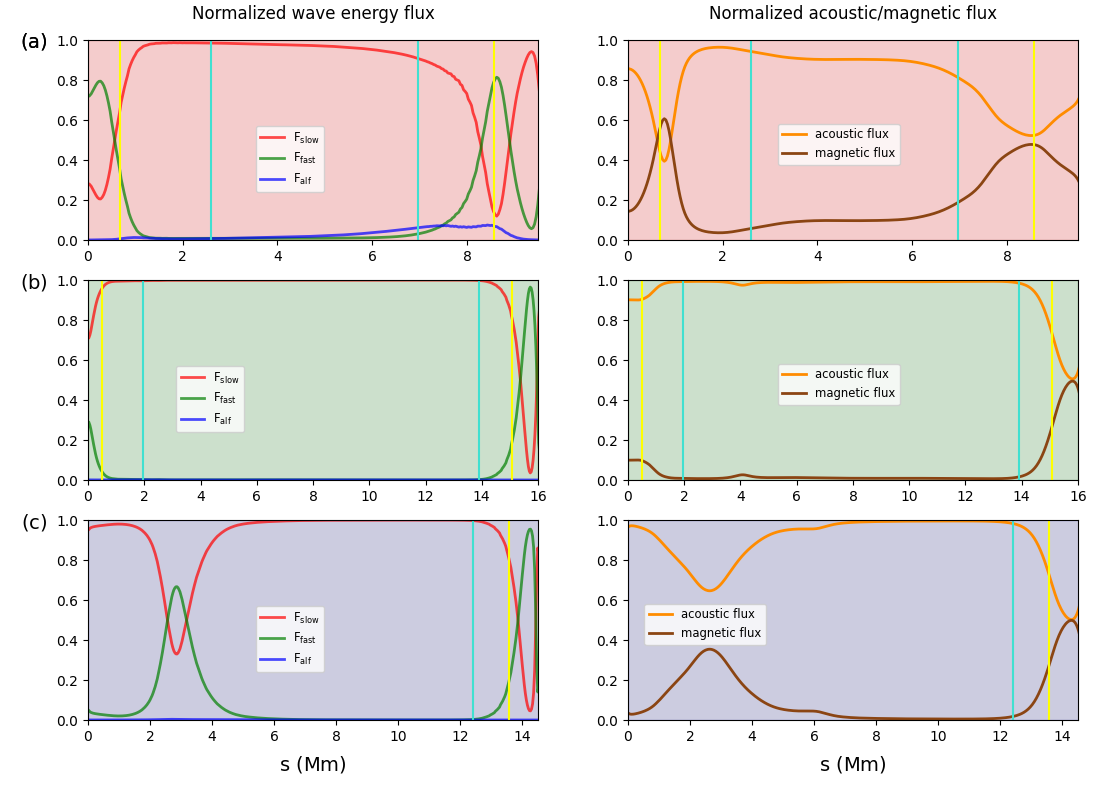}
         \caption{Energy flux comparisons for the three selected flux surfaces. The surfaces are indicated as background colors: red (top row), green (middle row), and blue (bottom row). Left column: Energy fluxes associated with slow, fast, and Alfv\'en waves. Right column: Acoustic and magnetic fluxes. The vertical lines represent the locations of the Alfvén-acoustic equipartition layer (yellow) and the transition region (turquoise).}
         \label{energy_flux}
\end{figure*}
\begin{figure*}
     \centering
        \includegraphics[scale=0.5]{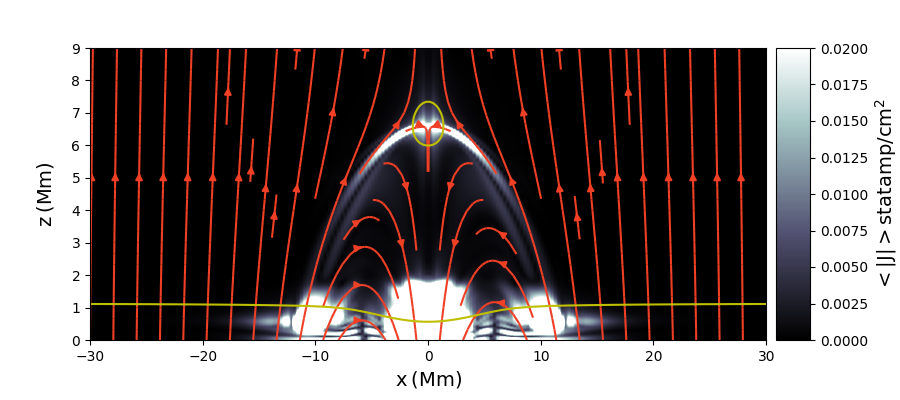}
         \caption{Spatial map of the magnitude of current (time-averaged over three wave periods) shown for a $y = 0$ slice through the domain.}
         \label{current}
\end{figure*}
Finally, we can use the decomposition to quantify the energy fluxes among the three MHD wave types, along the three previously discussed flux surfaces, using the method described in our earlier work. Here we only analyze the vertical driver case, the rotational counterpart is found in \citealt{2022A&A...660A..21Y}.
Fig. \ref{energy_flux} compares the wave energy flux distribution among three MHD wave modes (left column) as well as among acoustic and magnetic components (right column) for three selected constant flux surfaces (the surfaces are the ones shown in our current Fig.~\ref{one_instant_with_surface_lines}).
From the left column, it is clear that most of the wave energy flux is associated with the slow magneto-acoustic wave, which is expected as our wave driver is in vertical velocity, and is almost parallel to the background magnetic field in the layers close to the solar surface. Note that the green surface has the driver directly acting near the leftmost $s=0$ region.
There, the velocity perturbation is hence a longitudinal velocity vector corresponding to slow magneto-acoustic waves in low-plasma beta regions. These are indicated by the region in between the two yellow vertical lines in the top two rows of this figure. 
Since the blue surface passes very close to the null point, we get clear signatures of wave mode conversion there, seen in the obvious transformations happening at $s\approx 3$ Mm.
Energy flux available in fast waves increases rapidly, while energy flux available in slow waves decreases.
For all surfaces, the energy flux associated with Alfv\'en waves remains negligible compared to the other two wave modes.
Similarly, in the right column, we note that most of the energy flux is acoustic flux, which is expected as slow waves are of acoustic nature in the low plasma beta regions.
Again, in the case of the blue surface, the magnetic component of energy flux increases, and the acoustic component of energy flux decreases close to the null point, which can be attributed to mode conversion.

\subsection{Current accumulation at the null point}

Coronal null points are crucial from an energy perspective as these are the locations where the Alfv\'en speed is zero, and waves traveling with this speed can no longer propagate. 
Previous studies have shown that fast waves approaching the null point may wrap around it and lead to an increase in current density at the null point (\citealt{2004A&A...420.1129M, 2011Natur.475..477M}). 
To verify this current accumulation in our numerical setup, we calculated the time-averaged (over three wave periods) current density shown in Fig. \ref{current} for a vertical plane $y=0$, where the field lines are indicated in red. 
As expected, this vertical driver now indeed leads to strong current accumulation around the null point, which was absent in the case of the spiral driver (please see Fig. 9 of \citealt{2022A&A...660A..21Y} for comparison).
Considering we did not include resistivity to allow a fair comparison to our previous setup used in \citealt{2022A&A...660A..21Y} the current buildup around the magnetic null point suggests the potential for Joule dissipation and subsequent heating. 
We note that current accumulation near the region where the driver is located is mostly due to fast wave reflection happening there. 
Whereas the current accumulation at around $\pm$ 10 Mm is a boundary condition effect as the fast waves generated near the null point, travel and get reflected at the surface.

%%%%%%%%%%%%%%%%%%%%%%%%%%%%%%%%%%%%%%  
%%%%%%%%%%%%%%%%%%%%%%%%%%%%%%%%%%%%%%%%
\begin{figure}
     \centering
        \includegraphics[scale=0.6,trim=0 0 0 0]{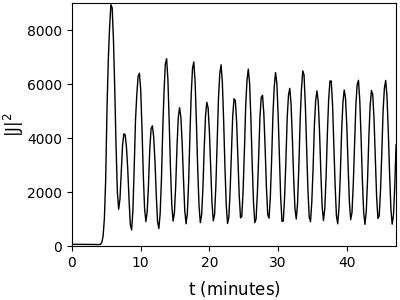}
             \caption{Time evolution of the square of the current density in the region surrounding the null point.}
             \label{qpp}
\end{figure}

Since the periodic wave driver causes the current to accumulate at the null point, we further looked into the potential heating signature and modulation in the thermal emission profile. 
We integrate the square of the absolute value of current in the surrounding region of the null point and plot its variation with time as shown in Fig. \ref{qpp}.
This quantity is directly proportional to Ohmic dissipation and gives an indirect estimate of temporal variation of plasma heating around the null point.
These periodic modulations in the magnitude of current around the null might result in periodic variations in plasma temperatures. They can be associated with quasi-periodic pulsations (QPPs) in solar and stellar flares or similar periodic energetic events.
Here, it is important to note that the oscillation frequency of the current magnitude is twice the frequency of the input wave driver. 
The possible relationship between p-modes and the periodicity of QPPs has been of great interest in the past few decades. 
There is observational evidence that the energy of 3-min oscillations in the active regions is significantly increased before the occurrence of the solar flare occurs and associated light curves show pronounced variations with similar periods and its higher harmonics (\citealt{2009A&A...505..791S,2019ApJ...875...33H,2015A&A...574A..53K}).

\section{Conclusions}\label{conclusion}
Magnetic null points are one manifestation of the complex topology of magnetic fields in the solar corona.
They are the locations in space where the magnetic field vanishes. They are regarded as proxies for magnetic reconnection sites and current dissipation. But due to their inherently three-dimensional nature, it is difficult to investigate the process of magnetic reconnection and current
dissipation around magnetic nulls using observations from one vantage point (\citealt{2019ApJ...874..157R, 2023NatCo..14.2107C}). Statistically, eruptive events are encountered more frequently in active
regions with null points than without null points (\citealt{2007ApJ...670L..53B}). Thus, it is crucial to investigate MHD dynamics using state-of-the-art simulations to unravel the physical mechanisms responsible for various processes associated with coronal magnetic null points.
To this end, we used our open-source \texttt{MPI-AMRVAC} code to perform simulations of the solar atmosphere ranging from the photosphere to the solar corona such that the atmosphere is gravitationally stratified and the magnetic field has a complex topology representative of a coronal null point. 

In observations, the periodicity of MHD waves is often shown to be linked with the periodicity of quasi-periodic oscillations detected in intensity enhancements at the null point (\citealt{2005A&A...440L..59F, 2011ApJ...740...90V, 2012ApJ...755..113S}).
\citet{2005LRSP....2....3N} attributed it to the influence of external evanescent or leaking parts of
the wave oscillation that can reach the null point.
Using MHD simulations \citet{2006SoPh..238..313C} demonstrated that five-minute solar p-mode oscillations can also lead to periodically triggered reconnection and thus transfer their periodicities in QPPs.
Whereas, \citet{2009A&A...494..329M} investigated the emergence of a buoyant flux tube in a stratified atmosphere permeated by a unipolar magnetic field and reported periodic reconnection taking place as the system searches for equilibrium.
More recently, \citet{2023ApJ...947L..17L} showed the formation of a null point as a result of the process of flux emergence and investigated the associated reconnecting plasmoid chains and the occurrence of multithermal jets.
\citet{2017ApJ...844....2T} demonstrated the interconnected nature of magnetic reconnection and MHD waves. They showed that magnetic reconnection at realistic 3D magnetic null points naturally proceeds in an oscillatory fashion, and produces MHD waves.
There are various other mechanisms proposed in the literature that can explain the observed periodicity of intensity enhancements at null points (e.g., see review by \citealt{2018SSRv..214...45M}). However, there is currently no consensus in the solar physics community on a physical mechanism that can unambiguously explain all quasi-periodic intensity variations at null points. In view of the fact that their periodicity does coincide in order of magnitude with MHD waves detected abundantly
in the solar corona, we examined the interaction of acoustic p-modes with coronal null topology and investigated wave transformations in detail.

To replicate the effect of global p-modes, we employed a vertical wave driver at the bottom boundary acting in the central region around the null spine. Since the background magnetic field is mostly vertical in the wave excitation region, most vertical velocity perturbation travels in slow waves. 
These slow waves undergo secondary wave mode conversion at the Alfv\'en acoustic equipartition layer, generating fast mode that leads to current accumulation at the null point. Fast modes generated at the first equipartition layer get reflected back from the sharp density gradients at the transition region and stay trapped below it.  
These results indirectly indicate that fast waves that are often observed at the null point have a different source of excitation rather than mode conversion taking place at the Alfv\'en acoustic equipartition layer below the transition layer.
We show that the fast modes get focused at the null point and lead to localized currents, but they are not generated below the transition region but in the surrounding surface of the null point where another Alfv\'en acoustic equipartition condition exists. 
Moreover, the periodicity of resistive dissipation is compatible with the second harmonic of wave driving, as has often been observed. Thus, we conjecture that although the periodicity of quasi-periodic intensity enhancement at the null point covers a vast range, it could possibly be determined by the dominant wave period excited at the solar surface. 
It is worthwhile to mention here that the current accumulation near the region where the driver is located is mostly due to fast wave reflection happening there. Whereas the current accumulation at around $\pm$ 10 Mm is a boundary condition effect as the fast waves generated near the null point, travel and get reflected at the surface.

Moreover, QPPs are often shown to exhibit periodicity consistent with higher harmonics of MHD waves (\citealt{2009McLaughlin, 2012A&A...544A..24T}). 
Analyzing the observations in X-ray, microwave, and extreme ultraviolet bands for the decay phase of an X-class solar flare \citet{2019MNRAS.483.5499K} reported that the second harmonic of the slow magneto-acoustic mode is the most likely explanation for the observed periods of QPPs.
However, they could not verify the possible mechanism behind the intensity enhancements at the periodicity corresponding to the second harmonic.
With the model presented here, we clearly demonstrate that mode conversion of p-modes may lead to the generation of fast modes surrounding the null point. We expect that the current fluctuations would produce intensity enhancement at the second harmonic by joule dissipation, which needs to be confirmed by follow-up resistive MHD simulations which can be used for synthetic views.
One limitation of the current model is that we dealt only with the linear regime of the MHD waves so that we could directly compare our results with our previous study (\citealt{2022A&A...660A..21Y}).
However, as a future step, it would be interesting to see how the results obtained here modify in the case of a nonlinear regime where we could also include the non-ideal terms in our setup such as thermal conduction and radiative losses.
\begin{acknowledgements}
NY and RK are supported by Internal funds KU Leuven, project C14/19/089 TRACESpace. RK further received funding from the European Research Council (ERC) under the European Union’s Horizon 2020 research and innovation programme (grant agreement no. 833251 PROMINENT ERC-ADG 2018), and FWO grant G0B4521N. NY acknowledges the support provided by DST INSPIRE Fellowship under grant number DST/INSPIRE/04/2021/001207. 
\end{acknowledgements}

% %%%%%%%%%%%%%%%%%%%%%%%%%%%%%%%%%%%%%%%%%%%%%%
 \bibliographystyle{aa} % stylefile aa.bst
 \bibliography{null} % Yourfile.bib
 %%%%%%%%%%%%%%%%%%%%%%%%%%%%%%%%%%%%%%%%%%
\end{document}